\documentclass[nofootinbib, aps,preprint]{revtex4}
\usepackage{eurosym}
\usepackage{amsfonts}
\usepackage{amsmath}
\usepackage{amssymb}
\usepackage{graphicx}
\usepackage{enumitem}
\providecommand{\U}[1]{\protect\rule{.1in}{.1in}}

\begin{document}

\preprint{}
\title{Frauchiger-Renner argument and quantum histories}
\author{Marcelo Losada}
\affiliation{Universidad de Buenos Aires and Università degli Studi di Cagliari}
\author{Roberto Laura}
\affiliation{Facultad de Ciencias Exactas, Ingenier\'{\i}a y Agrimensura (UNR) and
Instituto de F\'{\i}sica Rosario - CONICET }
\author{Olimpia Lombardi}
\affiliation{Universidad de Buenos Aires - CONICET}

\begin{abstract}
In this article we reconstruct the Frauchiger and Renner argument, taking into account that the assertions of the argument are made at different times. To do this, we use a formalism of quantum histories, namely the Theory of Consistent Histories. We show that the supposedly contradictory conclusion of the argument requires computing probabilities in a family of histories that does not satisfy the consistency condition, i.e., an invalid family of histories for the theory.
\end{abstract}

\date{October 2019}
\maketitle

\section{Introduction}

In April 2016, Frauchiger and Renner published an
article online in which they introduced a \textit{Gedankenexperiment} that led them
to conclude that \textquotedblleft no single-world interpretation can be
logically consistent\textquotedblright \, \cite{FyR2016}. In a new version of
the paper, the authors moderated their original claim, concluding \textquotedblleft that
quantum theory cannot be extrapolated to complex systems, at least not in a
straightforward manner\textquotedblright  \cite{FyR2018}.

Since its first online publication, the Frauchiger and Renner (F-R) argument
was extensively commented upon in the field of quantum foundations, since it was
considered as a new no-go result for quantum mechanics whose strength relies
on the fact that it is neutral regarding interpretation: on the basis of
three seemingly reasonable assumptions that do not include interpretive
premises, the argument leads to a contradiction. This fact has been
conceived as pointing to a deep shortcoming of quantum mechanics itself,
which contrasts with the extraordinary success of the theory.

In a previous article \cite{FyL2019} a careful reconstruction of the F-R
argument has been offered, which shows that the contradiction resulting from
the F-R argument is inferred by making classical conjunctions between
different and incompatible contexts, and, as a consequence, it is the result
of a theoretically illegitimate inference. However, recently\footnote{%
We thank Jeffrey Bub for pointing out this recent debate to us.} it has been
suggested that the criticism does not take into account the fact that the
inferences in the F-R argument are all carefully timed, and that this fact
would circumvent the objection based on the contextuality of quantum
mechanics. The purpose of this article is to analyze such a defense of the
F-R argument.

If timing really matters in the F-R argument, it seems natural to reconstruct the argument using a formalism of \textit{quantum histories}, which allows us to define logical operations between quantum properties at different times. The idea of quantum histories was mainly motivated by this limitation of quantum mechanics. In 1984, Robert Griffiths presented the first
version of his Theory of Consistent Histories \cite{Gri1984}; some years
later, he introduced some modifications to that original version \cite%
{Gri2002,Gri2013}. Roland Omn\`{e}s \cite%
{Omn1987,Omn1988a,Omn1988b,Omn1994,Omn1999} also published a series of
articles that contributed to the development of this theory. Simultaneously,
Murray Gell-Mann and James Hartle developed a similar formalism \cite%
{Hartle1991, GyH1993, GyH1990}. The Theory of Consistent Histories extends
the formalism of quantum mechanics. It introduces the notion of history, which generalizes the notion
of event: an elemental history is defined as a sequence of events at
different times, where an event is the occurrence of a property. But since
it is not possible to assign probabilities to the set of all histories, it
is necessary to select a subset of histories that satisfies additional
conditions.

In order to analyze the defense of the F-R argument on the basis of the fact
that the assertions are made at different times, we will carefully
reconstruct the argument in the framework of the Theory of Consistent
Histories. This task will allow us to prove that the supposedly
contradictory conclusion of the argument requires computing probabilities in
a family of histories that is not consistent, i.e., an invalid family of histories for the theory.

\section{The F-R argument}

\label{seccion FR}

The \textit{Gedankenexperiment} proposed in Frauchiger and Renner's article
is a sophisticated reformulation of Wigner's friend experiment \cite{Wigner1961}. In that original thought experiment, Wigner considers the
superposition state of a particle in a closed laboratory where his friend is
confined. When Wigner's friend measures the particle, the
state collapses to one of its components. However, from the outside of the
laboratory, Wigner still assigns a superposition state to the whole
composite system: \textit{Particle} $+$ \textit{Friend} $+$ \textit{Laboratory}.

\begin{figure}
	\includegraphics[scale=0.5]{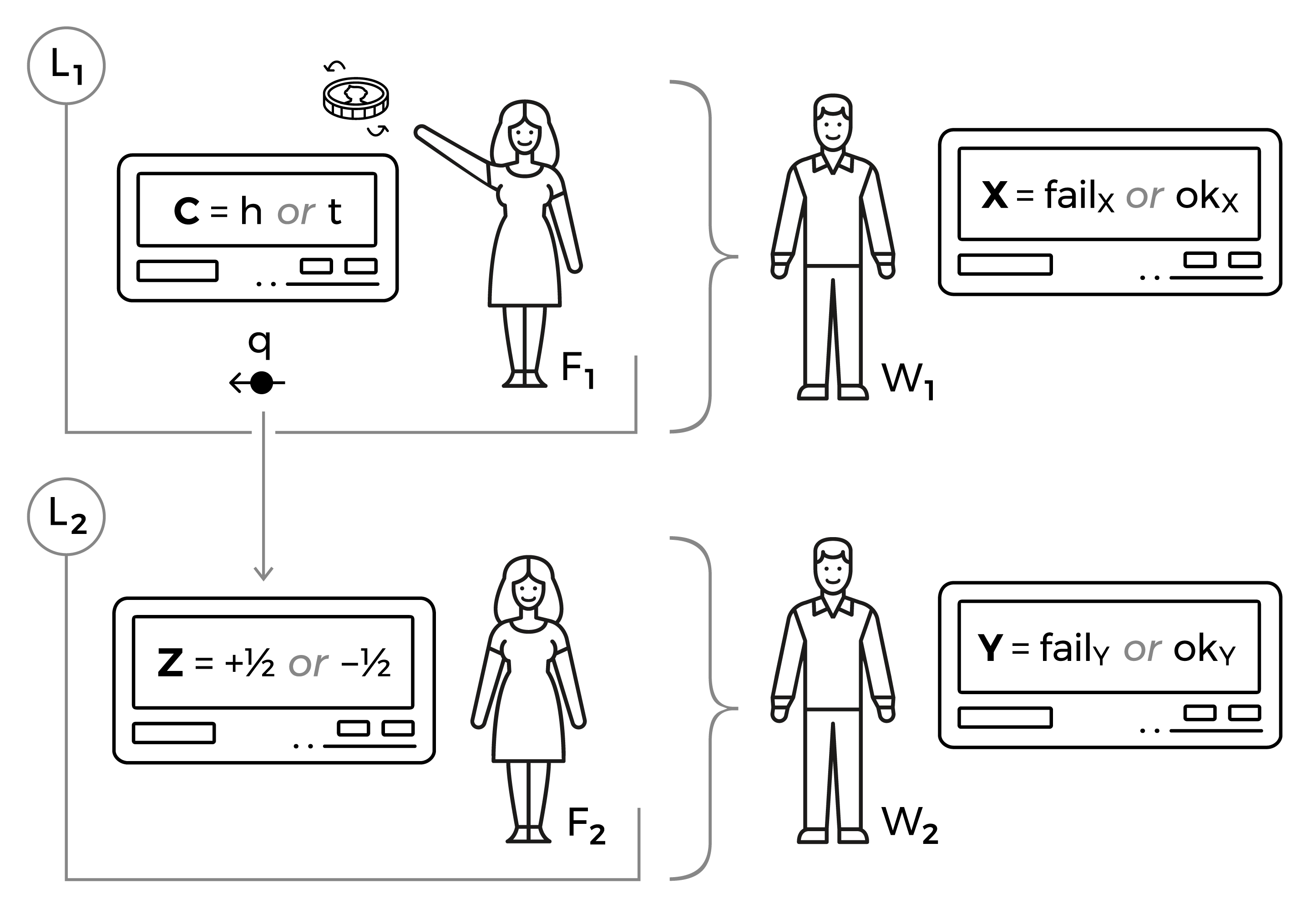}
	\caption{Illustration of the \textit{Gedankenexperiment}. Friend $F_1$ tosses a coin and measures its result. Depending on the outcome, she sends a qubit in a particular state. Then, Friend $F_2$ measures the spin of the qubit in the $z$ direction, obtaining $z= +\frac{1}{2}$ or $-\frac{1}{2}$. Finally, observers $W_1$ and $W_2$ measure the entire laboratories $L_1$ and $L_2$ obtaining outcomes fail$_X$ or ok$_X$ and fail$_Y$ or ok$_Y$, respectively.}
		\label{figura}
\end{figure}

The F-R argument relies on duplicating Wigner's setup (Fig. \ref{figura}). Let us consider two
friends $F_1$ and $F_2$ located in separate and isolated laboratories $L_1$ and $L_2$%
.\footnote{%
We slightly modify the original terminology for clarity.} $F_1$ measures the
observable $C$ of a biased ``quantum coin'' in the state $|\phi\rangle= 
\frac{1}{\sqrt{3}}|h\rangle+\sqrt{\frac{2}{3}}|t\rangle $, where $|h \rangle 
$ and $|t\rangle$ are the eigenstates of $C$, and $h$ and $t$ are its
respective eigenvalues. $F_1$ prepares a qubit in the state $|\downarrow
\rangle$ if the outcome is $h$, or in the state $|\rightarrow \rangle = 
\frac{|\uparrow \rangle +|\downarrow \rangle}{\sqrt{2}} $ if the outcome is $%
t$, and sends it to $F_2$. When $F_2$ receives the qubit, she measures its
observable $S_{z}$. After these two measurements, the state of the whole
system composed of the two laboratories is: 
\begin{equation}  \label{estado final1}
|\Psi\rangle = \frac{1}{\sqrt{3}} |H\rangle|\Downarrow \rangle +\sqrt{\frac{2%
}{3}} |T \rangle|\Rightarrow\rangle,
\end{equation}
where we have the following:

\begin{itemize}
\item $|H \rangle$ and $|T \rangle$, eigenstates of an observable $A$ with
eigenvalues $H$ and $T$, are the states of the entire laboratory $L_1$ when the
outcome of $F_1$'s measurement is $h$ and $t$, respectively.

\item $|\Uparrow \rangle$ and $|\Downarrow \rangle$, eigenstates of an
observable $B$ with eigenvalues $\Uparrow$ and $\Downarrow$, are the states
of the entire laboratory $L_2$ when the outcome of $F_2$'s measurement is $+1/2$
and $-1/2$, respectively.

\item $|\Rightarrow \rangle = \frac{1}{\sqrt{2}}|\Uparrow \rangle + \frac{1}{%
\sqrt{2}} ||\Downarrow \rangle$.
\end{itemize}

The \textit{Gedankenexperiment} continues by considering two ``Wigner''
observers, $W_1$ and $W_2$, located outside the laboratories, who will respectively
measure the observables $X$ and $Y$ of the laboratories $L_1$ and $L_2$, respectively:

\begin{itemize}
\item $X$ has the eigenvectors $|\text{fail}_{X} \rangle$ and $|\text{ok}%
_{X} \rangle$, such that: 
\begin{equation}
|\text{fail}_{X} \rangle = \frac{1}{\sqrt{2}}|H \rangle + \frac{1}{\sqrt{2}}
|T \rangle, ~~~~~ |\text{ok}_{X} \rangle= \frac{1}{\sqrt{2}}|H \rangle - 
\frac{1}{\sqrt{2}} |T \rangle,
\end{equation}

\item $Y$ has the eigenvectors $|\text{fail}_{Y} \rangle$ and $|\text{ok}%
_{Y} \rangle$, such that: 
\begin{equation}  \label{fail y}
|\text{fail}_{Y} \rangle = \frac{1}{\sqrt{2}}|\Downarrow \rangle + \frac{1}{%
\sqrt{2}} |\Uparrow \rangle, ~~~~~ |\text{ok}_{Y} \rangle = \frac{1}{\sqrt{2}%
}|\Downarrow \rangle - \frac{1}{\sqrt{2}} |\Uparrow \rangle.
\end{equation}
\end{itemize}

Before analyzing the consequences of the experiment, Frauchiger and Renner
point out that their argument can be conceived as a no-go theorem \cite%
{FyR2018} that proves that three ``natural-sounding'' assumptions, $(Q)$, $%
(C)$, and $(S)$, cannot all be valid simultaneously:\footnote{%
In the 2016 paper, Frauchiger and Renner implicitly consider $(Q)$ and $(C)$
as unavoidable: as a consequence, they claim that their argument shows that
``no single-world interpretation can be logically consistent'' and,
therefore, ``we are forced to give up the view that there is one single
reality'' \cite{FyR2016}. By contrast, in the 2018 paper, they stress that
``the theorem itself is neutral in the sense that it does not tell us which
of these three assumptions is wrong'' \cite{FyR2018}; as a consequence, they
admit the possibility of different theoretical and interpretive viewpoints
regarding their result, and include a table that shows which of the three
assumptions each interpretation of quantum theory violates.}
\begin{itemize}
	\item[$(Q)$]\textit{Compliance with quantum theory}: Quantum mechanics is universally valid, that is, it applies to systems of any complexity, including observers. Moreover, an agent knows that a given proposition is true whenever the Born rule assigns probability 1 to it.

	\item[$(C)$]\textit{Self-consistency}: Different agents' predictions are not contradictory.

	\item[$(S)$]\textit{Single-world}: From the viewpoint of an agent who carries out a particular measurement, this measurement has one single outcome.
\end{itemize}

On the basis of the above considerations --experimental setup and assumptions-- the F-R argument proceeds as follows. First, in order to
compute the probability that the measurements of $X$ and $Y$ yield the
results $|\text{ok}_{X}\rangle$ and $|\text{ok}_{Y}\rangle$, respectively,
the state described by equation \eqref{estado final1} must be expressed as: 
\begin{equation}  \label{estado final2}
|\Psi\rangle = \frac{1}{\sqrt{12}} |\text{ok}_{X}\rangle|\text{ok}%
_{Y}\rangle- \frac{1}{\sqrt{12}} |\text{ok}_{X}\rangle|\text{fail}%
_{Y}\rangle+\frac{1}{\sqrt{12}} |\text{fail}_{X}\rangle|\text{ok}_{Y}\rangle
+\sqrt{\frac{3}{4}} |\text{fail}_{X}\rangle|\text{fail}_{Y}\rangle.
\end{equation}
From this equation, it is clear that the probability of obtaining $\text{ok}%
_{X}$ and $\text{ok}_{Y}$ is $1/12$.

The second part of the argument consists in showing that the observers
involved in the experiment can draw a conclusion different from the above
one on the basis of the following reasoning. Let us consider the probability that $F_2$ obtains $-1/2$
in her $S_z$ measurement and $W_1$ obtains $|\text{ok}_X \rangle$ in her $X$
measurement; in order to compute this probability, the state described by
equation \eqref{estado final1} must be expressed as: 
\begin{equation}
|\Psi\rangle = \sqrt{\frac{2}{3}} |\text{fail}_{X}\rangle|\Downarrow \rangle
+ \frac{1}{\sqrt{6}} |\text{fail}_{X}\rangle|\Uparrow \rangle - \frac{1}{%
\sqrt{6}} |\text{ok}_{X}\rangle|\Uparrow\rangle.
\end{equation}
From this equation it can be inferred that such a probability is
zero. Then, if $W_1$ obtains $|\text{ok}_X\rangle$ in her $X$ measurement on the
laboratory $L_1$, she can infer with certainty that the outcome of $F_2$'s $S_z$ measurement on
the qubit was $+1/2$. In turn, if $F_2$ obtains $+1/2$ in her $S_z$
measurement on the qubit, she can infer that the outcome of $F_1$'s $C$
measurement on the quantum coin was $t$, because otherwise $F_1$ would send $F_2$ the qubit in state $|\downarrow\rangle$, see equation \eqref{estado final1}. And if $F_1$ obtains $t$ in
her $C$ measurement on the quantum coin, she can infer that the outcome of $%
W_2$'s $Y$ measurement on the laboratory $L_2$ will be $|\text{fail}_Y\rangle$, because
the outcome $t$ is perfectly correlated with the state $|\Rightarrow \rangle$
of the laboratory $L_2$, and $|\Rightarrow \rangle = |\text{fail}_Y \rangle$, see equation
\eqref{fail y}. Therefore, from a nested reasoning it can be concluded that,
when $W_1$ gets $|\text{ok}_X\rangle$, she can infer that $W_2$ certainly
gets $|\text{fail}_Y\rangle$. But this conclusion contradicts what was
inferred from equation \eqref{estado final2}, that is, that there is a non-zero
probability that $W_1$ gets $|\text{ok}_X\rangle$ and $W_2$ gets $|\text{ok}%
_Y\rangle$.

The reactions to the F-R argument have been multiple and varied (see for
example \cite{Sudbery2017, Bub2018, Brukner2018, Healey2018, Laloe2018,
LyH2018, Dieks2019}, just to mention some of them). However, since the
argument from which the contradiction is obtained involves quantum
properties at different times, it seems natural to consider a description of
the \textit{Gedankenexperiment} using the theory of quantum histories. This theory extends the formalism of quantum mechanics introducing the notion of quantum history: an elemental history is defined as a sequence of quantum properties at different times (see Sec. \ref{seccion historias}).
As far as we know, there has not been a detailed reconstruction of the argument in
terms of the Theory of Consistent Histories. In Section \ref{seccion
historias} we will offer such a description and we will draw the conclusions
that this formalism offers for this case.

Moreover, the vectors $|H\rangle$ and $|T\rangle$ of the previous discussion
are states of the measurement instrument in the laboratory $L_1$, while the vectors $%
|\Uparrow \rangle$ and $|\Downarrow \rangle$ are states of the measurement
instrument in the laboratory $L_2$. However, the states of the measurement instruments
of observers $W_1$ and $W_2$ are not included. In the next section we give a
complete description of the process including the Hilbert spaces
corresponding to all measurement instruments.

\section{The diachronic development of the argument}

\label{seccion3}

Let us recall that in the laboratory $L_1$ there is a quantum coin in the initial state 
\begin{equation}
|\phi\rangle= \frac{1}{\sqrt{3}}|h\rangle+\sqrt{\frac{2}{3}}|t\rangle \in 
\mathcal{H}_C ,
\end{equation}
where $\mathcal{H}_C$ is the Hilbert space of the coin. The initial state of
the rest of the laboratory $L_1$ (including observer $F_1$) is $|a_0\rangle \in 
\mathcal{H}_{F_1}$. Therefore, the Hilbert space of the entire laboratory $L_1$ is $%
\mathcal{H}_{L_1}= \mathcal{H}_{C} \otimes \mathcal{H}_{F_1}$. In turn, in
the laboratory $L_2$ there is a qubit, which initially is in state $|q_0\rangle \in 
\mathcal{H}_{q}$, where $\mathcal{H}_{q}$ is the Hilbert space of the qubit.
The initial state of the rest of laboratory $L_2$ (including observer $F_2$) is 
$|b_0\rangle \in \mathcal{H}_{F_2}$. Therefore, the Hilbert space of the
entire laboratory $L_2$ is $\mathcal{H}_{L_2}= \mathcal{H}_{q} \otimes \mathcal{H}%
_{F_2}$.

Observer $W_1$ measures the observable $X$ of the laboratory $L_1$ with an
apparatus, which is initially in a state $|w_{10}\rangle \in \mathcal{H}%
_{W_1}$, where $\mathcal{H}_{W_1}$ is the Hilbert space of the apparatus. In
turn, observer $W_2$ measures the observable $Y$ of the laboratory $L_2$ with an
apparatus initially in a state $|w_{20}\rangle \in \mathcal{H}_{W_2}$, where 
$\mathcal{H}_{W_2}$ is the Hilbert space of the corresponding apparatus.

Summing up, the Hilbert space of the entire process is $\mathcal{H}= 
\mathcal{H}_{L_1} \otimes \mathcal{H}_{L_2}\otimes \mathcal{H}_{W_1}\otimes 
\mathcal{H}_{W_2}$, and the initial state at time $t_0$ is 
\begin{equation}  \label{estado inicial}
|\Psi_0\rangle= |\phi\rangle \otimes | a_0\rangle \otimes |q_0 \rangle
\otimes|b_0\rangle \otimes|w_{10}\rangle \otimes|w_{20}\rangle \in \mathcal{H%
}.
\end{equation}

In what follows, we describe the consecutive processes.

\begin{itemize}
\item Time interval $(t_0, t_1)$: Observer $F_1$ measures the quantum coin.

This process is represented by a unitary evolution $U_{10}$ in the Hilbert
space $\mathcal{H}_{L_1}= \mathcal{H}_{C} \otimes \mathcal{H}_{F_1}$,
satisfying 
\begin{align}
U_{10}\left( |h\rangle \otimes |a_0 \rangle \right) = |h\rangle \otimes| a_h
\rangle \equiv |H \rangle , ~~~~ U_{10}\left( |t\rangle \otimes |a_0 \rangle
\right) = |t\rangle \otimes| a_t \rangle \equiv |T \rangle.
\end{align}

\item Time interval $(t_1, t_2)$: Observer $F_1$ prepares the qubit.

This process is represented by a unitary evolution $U_{21}$ in the Hilbert
space $\mathcal{H}_{F_1} \otimes \mathcal{H}_{q}$, satisfying 
\begin{align}
U_{21}\left( |a_h\rangle \otimes |q_0 \rangle \right) = |a_h\rangle \otimes|
\downarrow \rangle , ~~~~ U_{21}\left( |a_t\rangle \otimes |q_0 \rangle
\right) = |a_t\rangle \otimes|\rightarrow \rangle.
\end{align}

\item Time interval $(t_2, t_3)$: Observer $F_2$ measures the qubit

This process is represented by a unitary evolution $U_{32}$ in the Hilbert
space $\mathcal{H}_{L_2}=\mathcal{H}_{q} \otimes \mathcal{H}_{F_2}$,
satisfying 
\begin{align}
U_{32}(|\downarrow \rangle \otimes | b_0 \rangle) = |\downarrow \rangle
\otimes | b_\downarrow \rangle \equiv | \Downarrow \rangle , ~~~~
U_{32}\left( |\uparrow\rangle \otimes | b_0 \rangle \right) = |\uparrow
\rangle \otimes| b_\uparrow \rangle \equiv | \Uparrow \rangle.
\end{align}

\item Time interval $(t_3, t_4)$: Observer $W_1$ measures the laboratory $L_1$.

This process is represented by a unitary evolution $U_{43}$ in the Hilbert
space $\mathcal{H}_{L_{1}}\otimes \mathcal{H}_{W_{1}}$, satisfying 
\begin{equation*}
U_{43}\left( |\text{fail}_{X}\rangle \otimes |w_{10}\rangle \right) =|\text{%
fail}_{X}\rangle \otimes |w_{1\,\text{fail}}\rangle ,~~~~U_{43}\left( |\text{%
ok}_{X}\rangle \otimes |w_{10}\rangle \right) =|\text{ok}_{X}\rangle \otimes
|w_{1\,\text{ok}}\rangle .
\end{equation*}

\item Time interval $(t_4, t_5)$: Observer $W_2$ measures the laboratory $L_2$.

This process is represented by a unitary evolution $U_{54}$ in the Hilbert
space $\mathcal{H}_{L_{2}}\otimes \mathcal{H}_{W_{2}}$, satisfying 
\begin{equation*}
U_{54}\left( |\text{ok}_{Y}\rangle \otimes |w_{20}\rangle \right) =|\text{ok}%
_{Y}\rangle \otimes |w_{2\,\text{ok}}\rangle ,~~~~U_{54}\left( |\text{fail}%
_{Y}\rangle \otimes |w_{20}\rangle \right) =|\text{fail}_{Y}\rangle \otimes
|w_{2\,\text{fail}}\rangle .
\end{equation*}
\end{itemize}

Once the steps for the time evolution are stablished, the argument leading
to the contradictory result, reviewed in Section \ref{seccion FR}, should be
written in terms of probabilities involving properties at different times.
For example, in Section \ref{seccion FR} the value $1/12$ was obtained for
the probability for obtaining $\text{ok}_{X}$ and $\text{ok}_{Y}$.
Considering the description of the time evolution given in this section, we
should write 
\begin{equation}
\Pr \big( \left\lbrace  w_{2\,\text{ok}}\text{ at }t_{5}\right\rbrace  \wedge \left\lbrace  w_{1\,\text{ok}} \text{ at } t_{4}\right\rbrace \big)  =\frac{1}{12},  \label{primera parte}
\end{equation}%
where $\wedge $ represents the logical conjunction. This expression represents the probability for the measurement instrument of observer $%
W_{1}$ to indicate $w_{1\,\text{ok}}$ at time $t_{4}$ and for the measurement instrument of observer $W_{2}$ to indicate $w_{2\,\text{ok}}$ at the later time $t_{5}$.

The second part of the argument is based on the following conditional
probabilities 
\begin{align}
\Pr \big(  \left\lbrace b_{\uparrow} \text{ at } t_3 \right\rbrace \mid \left\lbrace  w_{1\, \text{ok}} \text{ at } t_4 \right\rbrace \big) &=1,  \label{primera} \\
\Pr \big(  \left\lbrace  a_t  \text{ at } t_1 \right\rbrace \mid \left\lbrace  b_{\uparrow}  \text{ at } t_3 \right\rbrace \big)%
&=1,  \label{segunda} \\
\Pr \big( \left\lbrace  w_{2\, \text{fail}}  \text{ at } t_5 \right\rbrace  \mid  \left\lbrace  a_t  \text{ at } t_1 \right\rbrace \big)%
&= 1.  \label{tercera}
\end{align}
If the last three conditional probabilities could be considered
simultaneously, then we could infer the following conditional probability:
\begin{align}
\Pr \big( \left\lbrace  w_{2\, \text{fail}} \text{ at } t_5 \right\rbrace  \mid \left\lbrace w_{1\, \text{ok}%
}  \text{ at } t_4  \right\rbrace \big) &=1.
\end{align}
Hence, $\Pr \big( \left\lbrace  w_{2\, \text{ok}} \text{ at } t_5 \right\rbrace  \land  \left\lbrace w_{1\, 
\text{ok}} \text{ at } t_4 \right\rbrace \big) =0$, which is in contradiction with equation %
\eqref{primera parte}.

Since the previous argument involves logical operations  between quantum properties at different times, it seems natural to analyze it using a formalism of quantum histories. In order to search for the possibility of obtaining equations \eqref{primera parte}, %
\eqref{primera}, \eqref{segunda} and \eqref{tercera} simultaneously, in the
next section we will apply the Theory of Consistent Histories.

\section{The F-R argument in terms of quantum histories}\label{seccion historias}

In what follows we present a brief summary of the Theory of Consistent
Histories (TQH) \cite%
{Gri1984,Gri2002,Gri2013,GyH1990,GyH1993,Hartle1991,Omn1987,Omn1988a,Omn1988b,Omn1994,Omn1999}. In quantum mechanics, the properties of a system are represented by
orthogonal projectors. Since an elementary history is a sequence of
properties at consecutive times, the TQH represents each elementary history
with a tensor product of orthogonal projectors. For example, a history of $n$ times $%
\breve{\Pi}= \Pi_1 \otimes ... \otimes \Pi_n$ represents a sequence of
properties $\Pi_1$, ..., $\Pi_n$, at times $t_1$, ..., $t_n$.

To define probabilities for quantum histories, it is necessary to
define a family of histories. For this purpose, first we have to choose a
context of properties at each time $t_i$, i.e., a set of
projectors that sum the identity of $\mathcal{H}$ and that are mutually
orthogonal: 
\begin{equation}
\Pi_{k_{i}}\Pi_{k_{i}^{\prime}}=\delta_{k_{i}k_{i}^{\prime}}\;\Pi_{k_{i}},%
\qquad {\textstyle\sum\nolimits_{k_{i}}} \Pi_{k_{i}}=I_\mathcal{H},\qquad
k_{i},k_{i}^{\prime}\in\sigma_{i},\qquad i=1,...,n;  \notag
\end{equation}
where $I_{\mathcal{H}}$ is the identity of the Hilbert space $\mathcal{H}$, and each $\sigma_i$ is an index set.

Second, we define the atomic histories $\breve{\Pi}_{k_{1}, ..., k_{n} }$,
choosing one projector $\Pi_{k_i}$ at each time $t_i$: 
\begin{equation}
\breve{\Pi}_{k_{1}, ..., k_{n} }= \Pi_{k_{1}}\otimes...\otimes
\Pi_{k_{n}},\qquad (k_{1},...,k_{n})\in \breve{\sigma}, \qquad \breve{\sigma}
= \sigma_1 \times ... \times \sigma_n.  \notag  \label{mona}
\end{equation}
Then, we define the histories $\breve{\Pi}_{\Lambda}$ summing the histories $%
\breve{\Pi}_{k_{1},...,k_{n}}$ with $(k_{1},...,k_{n}) \in \Lambda\subseteq 
\breve{\sigma}$, i.e., $\breve{\Pi}_{\Lambda}= {\textstyle%
\sum\nolimits_{(k_{1}, ..., k_{n} )\in\Lambda}}\breve{\Pi}_{k_{1}, ...,
k_{n}}$. These histories represent disjunctions of the histories $\breve{\Pi}%
_{k_{1},...,k_{n}}$. Finally, the family of histories is the set obtained by
making arbitrary disjunctions between product histories.

If $\rho _{0}$ is the initial state at time $t_{0}$, the probability of a
general history $\breve{\Pi}_{\Lambda }$ is defined in the following way: 
\begin{equation}
\text{Pr}_{\rho _{0}}(\breve{\Pi}_{\Lambda })=\text{Tr}\left[ C^{\dag }(\breve{\Pi%
}_{\Lambda })\rho _{0}C(\breve{\Pi}_{\Lambda })\right],  \label{prob general2}
\end{equation}%
where we have introduced the chain operator $C(\breve{\Pi}_{\Lambda })={%
\textstyle\sum\nolimits_{(k_{1},...,k_{n})\in \Lambda }}C(\breve{\Pi}%
_{k_{1},...,k_{n}})$, in which 
\begin{equation}
C(\breve{\Pi}_{k_{1},...,k_{n}})= U(t_{0},t_{1})\Pi _{k_{1}}U(t_{1},t_{2})\Pi
_{k_{2}} \, ... \, U(t_{n-1},t_{n})\Pi _{k_{n}}U(t_{n},t_{0}) \notag
\label{chain product}
\end{equation}
with $U(t_{i},t_{j})=e^{-iH(t_{i}-t_{j})/\hbar }$.

In general, the probability definition given in equation 
\eqref{prob
general2} does not satisfy the axiom of additivity. Therefore, to have a
well-defined probability, the atomic histories of a family of histories must satisfy an
additional condition, called the \textit{consistency condition}. 
\begin{equation}
\text{Tr}\left[ C^{\dag }(\breve{\Pi}_{k_{1},...,k_{n}})\rho _{0}C(\breve{\Pi%
}_{k_{1}^{\prime },...,k_{n}^{\prime }})\right] =0,~~~~\forall
\,(k_{1},...,k_{n})\neq (k_{1}^{\prime },...,k_{n}^{\prime }).  \label{weak}  
\end{equation}
Intuitively, the consistency condition measures the amount of interference between pairs of histories. When $n=1$, this condition is automatically satisfied, and the probability expression of equation \eqref{prob general2} reduces to the Born rule. However, in the general case, the consistency condition is not trivial, and when it is satisfied the probability expression provides a generalization of the Born rule.

In order to describe the F-R argument in terms of quantum histories, we
first obtain the probability for the measurement instrument of the observer $%
W_{1}$ to indicate $w_{1\,\text{ok}}$ at time $t_{4}$ and for the
measurement instrument of the observer $W_{2}$ to indicate $w_{2\,\text{ok}}$
at a later time $t_{5}$.

A suitable context of properties for time $t_{4}$ should include the
properties $w_{1\,\text{ok}}$, $w_{1\,\text{fail}}$ and it has to be completed
with the property $\lnot \left( w_{1\,\text{ok}}\vee w_{1\,\text{fail}%
}\right) $ (where $\vee $ is the disjunction and $\lnot $ is the negation) in order to include all the degrees of freedom of the measurement instrument, for example the initial state $|w_{1 0} \rangle$ given in equation \eqref{estado inicial}. These properties are represented by the following projectors: 
\begin{align}
\Pi _{w_{1\,\text{ok}}}& =I_{L_{1}}\otimes I_{L_{2}}\otimes |w_{1\,\text{ok}%
}\rangle \langle w_{1\,\text{ok}}|\otimes I_{W_{2}},  \notag \\
\Pi _{w_{1\,\text{fail}}}& =I_{L_{1}}\otimes I_{L_{2}}\otimes |w_{1\,\text{%
fail}}\rangle \langle w_{1\,\text{fail}}|\otimes I_{W_{2}},  \label{tiempo4}
\\
\Pi _{\lnot \left( w_{1\,\text{ok}}\vee w_{1\,\text{fail}}\right) }& =I_{%
\mathcal{H}}-\Pi _{w_{1\,\text{fail}}}-\Pi _{w_{1\,\text{ok}}},  \notag
\end{align}
where each $I_{K}$ is the identity of the corresponding Hilbert space $\mathcal{H}_{K}$. These three projectors provide a context of properties of the Hilbert space $%
\mathcal{H}$.

For time $t_{5}$, a suitable context of properties should include the
properties of the measurement instrument of observer $W_{2}$, i.e., $w_{1\,%
\text{ok}}$, $w_{1\,\text{fail}}$, and it has to be completed with the property $%
\lnot \left( w_{1\,\text{ok}}\vee w_{1\,\text{fail}}\right) $. These properties are represented by the following projectors: 
\begin{align}
\Pi _{w_{2\,\text{ok}}}& =I_{L_{1}}\otimes I_{L_{2}}\otimes I_{W_{1}}\otimes
|w_{2\,\text{ok}}\rangle \langle w_{2\,\text{ok}}|,  \notag \\
\Pi _{w_{2\,\text{fail}}}& =I_{L_{1}}\otimes I_{L_{2}}\otimes
I_{W_{1}}\otimes |w_{2\,\text{fail}}\rangle \langle w_{2\,\text{fail}}|,
\label{tiempo5} \\
\Pi _{\lnot \left( w_{2\,\text{ok}}\vee w_{2\,\text{fail}}\right) }& =I_{%
\mathcal{H}}-\Pi _{w_{2\,\text{fail}}}-\Pi _{w_{2\,\text{ok}}}.  \notag
\end{align}%
These three projectors also provide a context of properties of the Hilbert
space $\mathcal{H}$.

From the contexts of properties for times $t_{4}$ and $t_{5}$, we can
generate a family of two-times histories, whose atomic histories are $\breve{%
\Pi}_{k_{4},k_{5}}=\Pi _{k_{4}}\otimes \Pi _{k_{5}}$, with $\Pi _{k_{4}}$
one of the projectors of equations \eqref{tiempo4} and $\Pi _{k_{5}}$ one of
the projectors of equations \eqref{tiempo5}. It is easy to verify that the
family generated by these atomic histories satisfies the consistency
conditions given in equation \eqref{weak}. Therefore, equation 
\eqref{prob
general2} can be used to compute the probability of quantum history $\breve{%
\Pi}_{w_{1\,\text{ok}},w_{2\,\text{ok}}} = \Pi _{w_{1\,\text{ok}}} \otimes \Pi _{w_{2\,\text{ok}}}$,
\begin{equation}
\Pr \left( \breve{%
	\Pi}_{w_{1\,\text{ok}},w_{2\,\text{ok}}}\right)  =\frac{1}{12}.
\end{equation}
This shows that using the Theory of Consistent Histories, and explicitly
considering the measurement instruments as quantum systems, we obtain the
same result given in Section \ref{seccion3} for the first part of the
argument.

In the same way, different consistent families of two-times histories can be
defined to express equations \eqref{primera}, \eqref{segunda} and %
\eqref{tercera}. However, if the three equations are going to be used
together in the same argument, it is necessary to have a consistent family
of four-times histories including the possible results of the instrument of
the observer $F_1$ at time $t_1$, of the instrument of
the observer $F_2$ at time $t_3$, of the instrument of
observer $W_1$ at time $t_4$, and of the instrument of
observer $W_2$ at time $t_5$. 

For times $t_{4}$ and $t_5$, the contexts of properties given in equations \eqref{tiempo4} and \eqref{tiempo5} are adequate.
For time $t_{1}$, a suitable context of properties should include the
properties $a_{h}$, $a_{t}$, and it has to be completed
with the property $\lnot \left( a_{h}\vee a_{t}\right)$.
These properties are represented by the following projectors: 
\begin{align}
\Pi _{a_{h}}& = I_C \otimes |a_h \rangle \langle a_h |\otimes I_{L_{2}}\otimes I_{W_{1}}\otimes I_{W_{2}},  \notag \\
\Pi _{a_{t}}& =I_{C}\otimes |a_t \rangle \langle a_t |\otimes I_{L_{2}}\otimes I_{W_{1}} \otimes I_{W_{2}},  \label{tiempo1}
\\
\Pi _{\lnot \left( a_{h}\vee a_{t}\right) }& =I_{%
	\mathcal{H}}-\Pi _{a_{h}}-\Pi _{a_{t}},  \notag
\end{align}
For time $t_{3}$, a suitable context of properties should include the
properties of the measurement instrument of observer $F_{2}$, i.e., $b_{\downarrow}$, $b_{\uparrow}$, and it has to be completed with the property $
\lnot \left( b_{\downarrow}\vee b_{\uparrow}\right) $. These
properties are represented by the following projectors: 
\begin{align}
\Pi _{b_{\downarrow}}& =I_{L_{1}}\otimes I_{q}\otimes |b_{\downarrow} \rangle \langle b_{\downarrow} | \otimes  I_{W_{1}}\otimes I_{W_{2}},  \notag \\
\Pi _{b_{\uparrow}}& =I_{L_{1}}\otimes I_{q}\otimes |b_{\uparrow} \rangle \langle b_{\uparrow} | \otimes
I_{W_{1}}\otimes I_{W_{2}},
\label{tiempo3} \\
\Pi _{\lnot \left( b_{\downarrow}\vee b_{\uparrow}\right) }& =I_{%
	\mathcal{H}}-\Pi _{b_{\downarrow}}-\Pi _{b_{\uparrow}}.  \notag
\end{align}%

From the contexts of properties for times $t_1$, $t_3$, $t_{4}$ and $t_{5}$ we can
generate a family of four-times histories, whose atomic histories are 
\begin{equation}\label{atomic}
\breve{\Pi}_{k_1, k_3, k_4, k_5} =\Pi _{k_{1}} \otimes  \Pi _{k_{3}} \otimes  \Pi _{k_{4}} \otimes \Pi _{k_{5}},  
\end{equation}
 with  $\Pi _{k_{1}}$, $\Pi _{k_{3}}$, $\Pi _{k_{4}}$ and $\Pi _{k_{5}}$ 
projectors chosen from equations \eqref{tiempo1}, \eqref{tiempo3}, \eqref{tiempo4} and \eqref{tiempo5}, respectively.

The non-atomic histories that are involved in the F-R argument are the following:
\begin{align}
\breve{\Pi}_{1 \,t} &=\Pi_{a_t} \otimes  I_{\mathcal{H}}\otimes  I_{\mathcal{H}}\otimes I_{\mathcal{H}}, \\
\breve{\Pi}_{3\, \uparrow} &=I_{\mathcal{H}}\otimes   \Pi_{b_{\uparrow}}  \otimes I_{\mathcal{H}}\otimes I_{\mathcal{H}}, \\
\breve{\Pi}_{4\, \text{ok}} &= I_{\mathcal{H}}\otimes   I_{\mathcal{H}} \otimes \Pi _{w_{1\,\text{ok}}} \otimes I_{\mathcal{H}}, \\
\breve{\Pi}_{5 \, \text{fail}} &= I_{\mathcal{H}} \otimes  I_{\mathcal{H}}\otimes I_{\mathcal{H} } \otimes \Pi _{w_{2\,\text{fail}}}.
\end{align}
In terms of quantum histories, the F-R argument can be formulated as follows: 

 \textbf{First part}
\begin{equation}\label{parte1}
\Pr(\breve{\Pi}_{5 \, \text{ok}}  \land \breve{\Pi}_{4\, \text{ok}}) =\frac{1}{12}.
\end{equation}

\textbf{Second part}
 
\begin{equation}\label{premisas}
\Pr(\breve{\Pi}_{3\, \uparrow}|\breve{\Pi}_{4\, \text{ok}})=1, ~ \Pr(\breve{\Pi}_{1 \,t}|\breve{\Pi}_{3\, \uparrow})=1 ~ \text{and}  ~  \Pr(\breve{\Pi}_{5 \, \text{fail}} |\breve{\Pi}_{1 \,t})=1.
\end{equation}
This implies $\Pr(\breve{\Pi}_{5 \, \text{fail}} |\breve{\Pi}_{4 \, \text{ok}})=1$, and then $\Pr(\breve{\Pi}_{5 \, \text{ok}} |\breve{\Pi}_{4 \, \text{ok}})=0$. Therefore,
\begin{equation}\label{conclusion}
\Pr(\breve{\Pi}_{5 \, \text{ok}}  \land \breve{\Pi}_{4\, \text{ok}}) =0.
\end{equation}
The contradiction is obtained from equations \eqref{parte1} and \eqref{conclusion}.

In order to infer equation \eqref{conclusion} from the equations \eqref{premisas}, the quantum histories must belong to a single consistent family of histories, generated by the atomic histories of equations \eqref{atomic}. But such a family of histories does not
satisfy the consistency conditions, given in equation \eqref{weak}.

To prove this statement, let us consider two atomic histories $\breve{\Pi}_{a_t, b_{\uparrow}, w_{1\,\text{ok}}, w_{2\,\text{ok}}}$ and $\breve{\Pi}_{a_h, b_{\downarrow}, w_{1\,\text{ok}}, w_{2\,\text{ok}}}$, representing different results for the four
measurements, 
\begin{equation}
\breve{\Pi}_{a_t, b_{\uparrow}, w_{1\,\text{ok}}, w_{2\,\text{ok}}} =\Pi _{a_t} \otimes  \Pi _{b_{\uparrow}} \otimes  \Pi _{w_{1\,\text{ok}}} \otimes \Pi _{w_{2\,\text{ok}}},  
\end{equation}
\begin{equation}
\breve{\Pi}_{a_h, b_{\downarrow}, w_{1\,\text{ok}}, w_{2\,\text{ok}}} =\Pi _{a_h} \otimes  \Pi _{b_{\downarrow}} \otimes  \Pi _{w_{1\,\text{ok}}} \otimes \Pi _{w_{2\,\text{ok}}},  
\end{equation}
and with the following chain operators 
\begin{align}
C(\breve{\Pi}_{a_t, b_{\uparrow}, w_{1\,\text{ok}}, w_{2\,\text{ok}}})& =U(t_{0},t_{1})\Pi _{a_{t}}U(t_{1},t_{3})\Pi
_{b_{\uparrow }}U(t_{3},t_{4})\Pi _{w_{1\,\text{ok}}}U(t_{4},t_{5})\Pi
_{w_{2\,\text{ok}}}U(t_{5},t_{0}), \\
C(\breve{\Pi}_{a_h, b_{\downarrow}, w_{1\,\text{ok}}, w_{2\,\text{ok}}})& =U(t_{0},t_{1})\Pi _{a_{h}}U(t_{1},t_{3})\Pi
_{b_{\downarrow }}U(t_{3},t_{4})\Pi _{w_{1\,\text{ok}}}U(t_{4},t_{5})\Pi
_{w_{2\,\text{ok}}}U(t_{5},t_{0}).
\end{align}%

Considering the unitary time evolution of the complete quantum system and
the initial state defined in equation \eqref{estado inicial}, we obtain 
\begin{align}
C^{\dag }&(\breve{\Pi}_{a_t, b_{\uparrow}, w_{1\,\text{ok}}, w_{2\,\text{ok}}})|\Psi _{0}\rangle =C^{\dag }(\breve{\Pi}_{a_h, b_{\downarrow}, w_{1\,\text{ok}}, w_{2\,\text{ok}}})|\Psi _{0}\rangle = \nonumber \\ &=\frac{1}{\sqrt{12}}U(t_{0},t_{5})(|\text{ok}%
_{X}\rangle \otimes |\text{ok}_{Y}\rangle \otimes |w_{1\,\text{ok}}\rangle
\otimes |w_{2\,\text{ok}}\rangle),
\end{align}%
and therefore, according to equation \eqref{weak}, the consistency
condition gives 
\begin{align}
\text{Tr}&\left[ C^{\dag }(\breve{\Pi}_{a_t, b_{\uparrow}, w_{1\,\text{ok}}, w_{2\,\text{ok}}})|\Psi _{0}\rangle \langle
\Psi _{0}|C(\breve{\Pi}_{a_h, b_{\downarrow}, w_{1\,\text{ok}}, w_{2\,\text{ok}}})\right] = \nonumber \\
&=\langle \Psi _{0}|C(\breve{\Pi}_{a_h, b_{\downarrow}, w_{1\,\text{ok}}, w_{2\,\text{ok}}})C^{\dag }(\breve{\Pi}_{a_t, b_{\uparrow}, w_{1\,\text{ok}}, w_{2\,\text{ok}}})|\Psi _{0}\rangle =\frac{1}{12}%
\neq 0.
\end{align}%
This proves that the atomic histories $\breve{\Pi}_{a_t, b_{\uparrow}, w_{1\,\text{ok}}, w_{2\,\text{ok}}}$ and $\breve{\Pi}_{a_h, b_{\downarrow}, w_{1\,\text{ok}}, w_{2\,\text{ok}}}$ do not satisfy the consistency condition and, therefore, there is
no family of consistent histories to describe the results of the four
measurement instruments of the F-R experiment.
For this reason, the conclusion of the second part of the F-R argument cannot be asserted.

Summing up, since the conclusion of the second part of the F-R argument is
based on an illegitimate inference, the supposed contradiction of the F-R
argument does not hold.

\section{Conclusions}

In a previous article \cite{FyL2019} one of us argued that the contradiction
resulting from the F-R argument is inferred by making classical conjunctions
between different and incompatible contexts, and, as a consequence, it is the
result of a theoretically illegitimate inference. However, it has been
suggested that the criticism does not take into account the fact that the
inferences in the F-R argument are all carefully timed, and this fact would
circumvent the objection based on the contextuality of quantum mechanics. 

If timing really matters in the F-R argument, it seems natural to reconstruct it using a theory of quantum histories, a formalism that allows us to deal with quantum properties at different times. We applied the Theory of Consistent Histories, and we showed that the
contradiction resulting from the F-R argument is inferred by computing probabilities in a family of histories that is not consistent, i.e. an invalid family of histories for the theory.

\section*{ACKNOWLEDGMENTS}

This work is partially supported by the project ``For a semantic extension of the Quantum Computation Logic: theoretical aspects and possible implementations", funded by RAS [project code: RASSR40341]. We would like to thank Griselda Losada for the illustration (Fig. \ref{figura}). We thank Jeffrey Bub for pointing out a recent debate regarding the FR argument, and for suggesting a clear explanation of part of the argument.

\end{document}